\documentclass{elsart}

\usepackage{amssymb}

\usepackage{graphicx}
\usepackage[figuresright]{rotating}

\newcommand \tie {{\it i.e.}}

\newcommand \ra  {\rightarrow}

\newcommand \vp {\vec{p}}

 \newcommand \ep {\epsilon}

\newcommand \hf {\frac{1}{2}}
\newcommand \A {\alpha}

\newcommand \lc {\langle}
\newcommand \rc {\rangle}
\newcommand \prt {\partial}
\newcommand \D {\Delta}
\newcommand \sg {\sigma}

\newcommand \nt {\noindent}

\newcommand \bvec{\left( \begin{array}{c} }
\newcommand \evec{\end{array} \right)}

\newcommand \bea{\begin{eqnarray} }
\newcommand \eea{\end{eqnarray} }
\newcommand \nn {\nonumber}
\newcommand {\be} {\begin{equation}}
\newcommand {\ee} {\end{equation}}

\newcommand {\mbx} {\mbox{}}

\begin{document}

\begin{frontmatter}



\title{Deciphering the properties of hot and dense matter with hadron-hadron correlations}


\author{A.~Majumder}

\address{Department of physics, Duke University, Box 90305, Durham, NC 27708.}

\begin{abstract}
Two classes of jet correlations in hot and dense matter are explored.  
Correlations between very high transverse momentum hadrons within a jet 
sample the gluon density of the medium, where, 
the minimal modification on the same side as the trigger is consistent with the 
picture and parameters of partonic energy loss.
Lower momentum partons, sampled through softer correlations, due to their larger wavelengths 
are sensitive to the presence of 
composite structures in the medium. Scattering off such states may modify the dispersion relation of 
the radiated gluons resulting in conical patterns in the detected correlations.
\end{abstract}

\end{frontmatter}


\section{Introduction}
Ultra-relativistic collisions of heavy nuclei lead to the production of highly excited 
strongly interacting matter~\cite{Jacobs:2004qv}. Present day collisions  at the Relativistic Heavy-Ion 
Collider (RHIC) have yielded numerous startling observables; all of which point 
to the production of matter which is decidedly non-hadronic and possibly deconfined in nature~\cite{RHIC_Whitepapers}.
One of the primary signatures for non-hadronic behaviour
 is the collective nature of the matter, different from 
that observed in excited hadronic media created at lower energy colliders. 
Yet another observable is the opacity of the produced matter to the passage of hard 
jets through it~\cite{highpt}. It is indeed the large observed suppression of the single 
inclusive spectrum of high transverse momentum ($p_T$) hadrons that has yielded the estimate that the 
produced matter is 30 times as dense as normal nuclear matter. Such estimates are
based on the picture of partonic energy loss. 
In the first part of these proceedings, this mechanism is tested via the modification of the correlation 
between two high $p_T$ hadrons within the same jet. 
Very high momentum partons 
and radiated hard gluons~\cite{Gyulassy:2003mc}, selectively sampled 
through the observation of  high transverse momentum hadrons  (referred to as hard-hard correlations)~\cite{Adler:2002tq,Adams:2006yt},
tend to sample the partonic substructure of the medium. As a result, any formalism that reproduces 
the single inclusive spectrum must also be able to estimate the modified correlation 
between hadrons produced in the fragmentation of a single jet.

The observed collective behaviour of the produced matter, 
manifested in its radial and elliptic flow is, oddly enough, different from what 
would have been expected from perturbative QCD calculations, assuming a quasi-particle 
picture of the quark gluon plasma (QGP). This raises the question as to what are the 
prevalent degrees of freedom in the produced matter. There exist conjectures that 
such matter is composed of a tower of bound states of quasi-particle quarks and 
gluons~\cite{Shuryak:2004tx}.  In the second part of these proceedings we explore the 
possibility that softer gluons radiated from hard partons, 
due to their longer wavelengths, are influenced by such states in the medium. This  
modifies the dispersion relation of the gluons~\cite{Koch:2005sx,Majumder:2005sw,Casalderrey-Solana:2006sq,Ruppert:2005uz} 
leading to non-jet-like correlations between a high momentum hadron and a softer hadron 
(referred to as hard-soft correlations)~\cite{star-cone,Adler:2005ee}. 

\section{Hard-Hard correlations}

We commence with a study of 
the correlations between two hard particles on the same side. 
Such two-hadron correlations have been measured
both in Deep-Inelastic Scattering~(DIS) off large nuclei~\cite{dinezza04} and in high-energy heavy-ion
collisions~\cite{Adler:2002tq}. 
Within the energy ranges and angles explored,
it is most 
likely that both these particles have their origin 
in a single jet which is modified by its interaction 
with the medium. Hence, such analysis requires 
the introduction of a dihadron fragmentation function~($D_q^{h_1 h_2}$)~\cite{Majumder:2004wh}, 
which accounts for the number of pairs of particles fragmenting from a jet. 
In the case of modification in cold nuclear matter,  a nucleus $A$ with forward momentum $Ap^+$, 
and a quark structure function $f_q^A(x_B,Q^2)$ is struck with a virtual photon with four-momentum
$q \equiv [-Q^2/2q^-,q^-,\vec{0}_\perp]$. 
The modified dihadron fragmentation function, for the production of two 
hadrons~($h_1,h_2$, off the struck quark) with momentum fractions $z_1,z_2$, is calculated as,
\bea
\!\!\!\!\!\!\!\!\tilde{D}_q^{h_1,h_2} (z_1,z_2) &=& D_q^{h_1,h_2}(z_1,z_2) +
\int_0^{Q^2} \frac{dl_{\perp}^2}{l_{\perp}^2} \frac{\A_s}{2\pi} \left[ \int_{z_1+z_2}^1 \frac{dy}{y^2}
\left\{ \D P_{q\ra q g} (y,x_B,x_L,l_\perp^2) \right. \right.
\label{eq-dihdr-mod} \\
& &\hspace{-1in} \times \left.
D_q^{h_1,h_2} \left(\frac{z_1}{y},\frac{z_2}{y} \right)  +  \D P_{q\ra g q} (y,x_B,x_L,l_\perp^2)
D_g^{h_1,h_2} \left(\frac{z_1}{y},\frac{z_2}{y} \right) \right\} \nn \\
& & \hspace{-1in} \left. +\int_{z_1}^{1-z_2} \frac{dy}{y(1-y)}
\D \hat{P}_{q\ra q g} (y,x_B,x_L,l_\perp^2) D_q^{h_1}
\left(\frac{z_1}{y})\right) D_g^{h_2}\left(\frac{z_2}{1-y} \right)
+ (h_1 \ra h_2) \right] . \nn 
\eea
In the above, $x_B= -Q^2/2p^+q^-$, $x_L = l_\perp^2/2p^+q^-y(1-y)$,
$l_\perp$ is the transverse momentum of the radiated gluon, 
$\D P_{q\ra qg}$ and $\D P_{q\ra g q}$
are the modified splitting functions with momentum fraction $y$, 
whose forms are identical
to that in the modified single hadron fragmentation functions~\cite{guowang}.
The switch $(h_1 \ra h_2)$ is only meant for the last term, which
represents independent fragmentation of the quark and gluon after
the induced bremsstrahlung. The corresponding modified splitting
function is,
\bea
\D \hat{P}_{q\ra gq} = \frac{1+y^2}{1-y}
\frac{C_A 2\pi \A_s T^A_{qg} (x_B,x_L)}{(l_\perp^2+\lc k_\perp^2\rc)
N_c f_q^A(x_B,Q^2)}, \label{splitting_func}
\eea
($\D P_{q\ra qg}$ is similar in form but also contains
contributions from virtual corrections).
In the above, $C_A=3$, $N_c=3$, and $\lc k_\perp^2\rc$ is the
average intrinsic parton transverse momentum inside the nucleus.
Note that both modified splitting functions depend on the
quark-gluon correlation function $T^A_{qg}$ in the nucleus~(see Refs.~\cite{Majumder:2004pt} for 
detailed expressions) which
also determines the modification of single hadron fragmentation
functions.
As a result, no additional parameters are required.
Due to the existence of sum rules connecting  dihadron fragmentation
functions to single hadron fragmentation functions~($D_q^{h}$)~\cite{Majumder:2004wh}, 
one studies the modification of the conditional distribution for the second rank hadrons,
$
R_{2h}(z_2)\equiv \int d z_1 D_q^{h_1,h_2}(z_1,z_2) \Bigg/ \int d z_1 D_q^{h_1}(z_1).
\label{eq-corr}
$

In high-energy heavy-ion (or $p+p$ and $p+A$) collisions, jets are
always produced in back-to-back pairs. Correlations of two
high-$p_T$ hadrons in azimuthal angle generally have two Gaussian
peaks~\cite{Adler:2002tq,Adams:2006yt}. 
The integral of the near-side peak (after background
subtraction) over the azimuthal angle 
can be related to the associated hadron distribution or
the ratio of dihadron to single hadron fragmentation functions. 
In these experiments, one usually considers the integrated  
yield of  hadrons [over a range of transverse momentum ($p_T^{\rm assoc}$] associated 
with a trigger hadron of  higher momentum ($p_T^{\rm trig}$) \tie,

\bea
R_{2h} = \int dp_T^{\rm assoc}  \frac{d \sigma^{AB}}{dy d {p_T^{\rm trig}} d {p_T^{\rm assoc}} }
\mbox{\Huge{/}}  \frac{d \sigma^{AB}} {dy d {p_T^{\rm trig}}  }.
\eea
Estimations of the 
associated yields require a  calculation where two hadrons with a given $p_T^{\rm trig}$ and 
$p_T^{\rm assoc}$ will originate from a wide range of initial jet energies weighted by the initial 
hard cross sections (obtained via a convolution of the initial structure functions and hard 
partonic cross sections~\cite{Majumder:2004pt,Majumder:2006we}). Assuming 
factorization of initial and final state effects, the differential cross-section 
for the production of two high $p_T$ hadrons at midrapidity from the impact of 
two nuclei $A$ and $B$ at an impact parameter between $b_{min},b_{max}$ is given as, 

\bea
\frac{d \sigma^{AB}}{dy d {p_T^{\rm trig}} d {p_T^{\rm assoc}} } 
 &=&  \int_{\mbx_{b_{min}}}^{\mbx^{b_{max}}}\!\!\!d^2 b 
\int d^2 r t_A(\vec{r}+\vec{b}/2) t_B(\vec{r} - \vec{b}/2)  \label{AA_sigma}  \\
&\times& \!\! 2K \!\! \int \!\! d x_a d x_b  G^A_a(x_a,Q^2)  G^B_b(x_b,Q^2)  
 \! \frac{  d \hat{\sg}_{ab \ra cd} }{ d \hat{t}}\tilde{D}_c^h(z_1,z_2,Q^2),  \nn
\eea
\nt
where, $G^A_a(x_a,Q^2)  G^B_b(x_b,Q^2)$ represent the nuclear parton distribution functions to find a 
parton $a (b)$ with momentum fractions $x_a(x_b)$ in a nucleus $A(B)$, $t_A$ (and $t_B$) represents the 
nuclear thickness function and $d \hat{\sg}_{ab \ra cd} / d \hat{t}$ represents the hard 
parton cross section with the Mandelstam variable $\hat{t}$. 
The factor $K \simeq 2$ accounts for higher order corrections and is identical to that used for the 
single hadron cross sections.  The medium modified dihadron 
fragmentation function may be expressed as in Eq.~(\ref{eq-dihdr-mod}), with the modified 
splitting functions generalized from Eq.~(\ref{splitting_func}) to the case of heavy-ion collisions.

\begin{figure}[htbp]
\hspace{1cm}
\resizebox{1.75in}{1.75in}{\includegraphics[1.in,1.in][6in,6in]{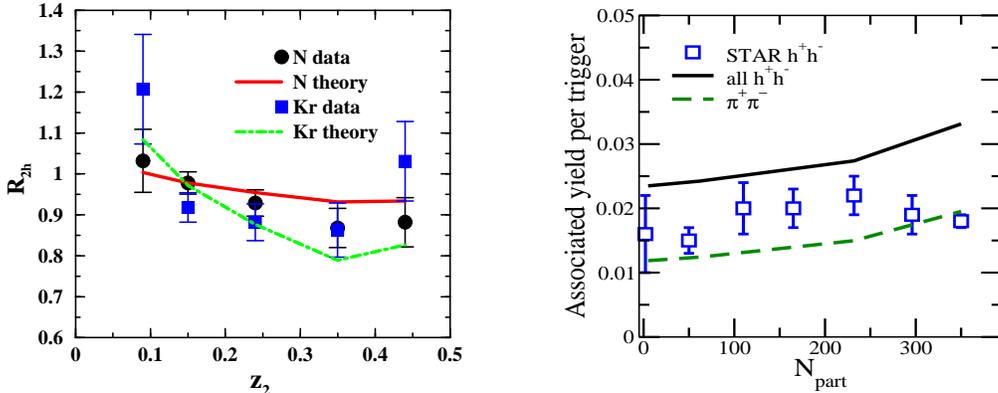}}
\hspace{2cm}
\resizebox{1.75in}{1.75in}{\includegraphics[0.2in,0.7in][4.5in,5.55in]{associated_yield.eps}}\caption{ Results of the medium modification of
    the associated hadron distribution in a cold nuclear medium 
    versus its momentum fraction (left panel) and versus system size in 
a hot medium (right panel) as compared to experimental data (see text for details).}
    \label{fig1}
\end{figure}

The associated hadron correlation
is found slightly suppressed in DIS off a nucleus versus
a nucleon target and moderately enhanced or unchanged in central $Au+Au$
collisions relative to that in $p+p$ (in sharp contrast
to the observed strong suppression of single inclusive
spectra in both DIS and central $A+A$ collisions~\cite{hermes1,highpt}).
Shown in the left panel of Fig.~\ref{fig1} is the predicted ratio of the associated
hadron distribution in DIS off a nuclear target ($N$ and $Kr$)
to that off a proton, as compared to the experimental
data from the HERMES collaboration at DESY~\cite{dinezza04}.
The suppression of the associated hadron distribution $R_{2h}(z_2)$
at large $z_2$ due to multiple scattering and induced gluon
bremsstrahlung in a nucleus is quite small compared to the
suppression of the single fragmentation functions~\cite{hermes1,EW1}.
The effect of energy loss seems to be borne, mainly, by the leading
hadron spectrum.

Computations of the associated yield in a heavy-ion collision, as a function of  the number of participants ($N_{part}$) 
are plotted in the right panel of  Fig.~\ref{fig1} along with 
experimental data from Ref.~\cite{Adams:2006yt}.  In this plot, $p_T^{\rm trig}$ ranges from $8-15$ GeV while 
the associated momentum  ranges from $6$ GeV $< p_T^{\rm assoc} < p_T^{\rm trig}$.  Unlike the case in 
DIS (left panel), the data are \emph{not} normalized by the associated yield in $p+p$ collisions. As a result, the 
normalization is sensitive to the flavour content of the detected hadrons. The experimental results include 
all charged hadrons (with certain decay corrections~\cite{Adams:2006yt}),
 whereas the theoretical predictions include two extreme possibilities: the lower dashed line 
denotes charged pions ($\pi^+,\pi^-$) inclusive of all decays and the upper solid line denotes 
$p,\bar{p},K^+,K^-$ and $\pi^+,\pi^-$ inclusive of all decays. The two cases bracket the experimental 
data, lending support to the framework of partonic energy loss. Decay corrections, which essentially involve  removing contributions from unstable particle decays to the detected flavour content,  will slightly reduce the plotted associated yields. 

While no trend may be discerned from 
the experimental measurements, 
the theoretical predictions show a slight enhancement with centrality due to increased 
trigger bias in more central collisions.
In central $Au+Au$ collisions, triggering on a high $p_T$ hadron 
biases toward a larger initial jet energy and therefore smaller 
$z_1$ and $z_2$. This is in contrast to the rise in the effective $z$ in the 
single inclusive estimates of Ref.~\cite{EW1}. In such observables 
there is no triggerring on the $p_T$ of the leading hadron and 
hence no bias of the initial parton energy.
This effect of trigger bias leads to an enhancement in the associated yield due to
the shape of dihadron fragmentation functions \cite{Majumder:2004wh}. 

An alternate explanation for the near constant associated yield as a function of 
centrality has been that of pure surface emission. In an extreme example of this 
view, one outlines a superficial annular ring in the transverse plane of a central 
heavy-ion collision. All partons produced on this outer annulus escape unmodified, 
while all partons produced in the region enclosed by the annulus are completely 
quenched and do not escape the dense system regardless of their energy. This 
view is somewhat naive, as higher momentum partons will naturally sample 
deeper into the medium. In the results presented above no such constraint on 
the production points has been assumed. 
The agreement between 
the theoretical predications and experimental results for the associated yield are thus 
highly non-trivial.

\section{Hard-Soft correlations}

For triggered events in central heavy-ion collisions, the Gaussian peak 
associated with the distribution of high $p_T$ associated particles on 
the away side is almost absent. As the $p_T$ of the associated 
particle is reduced, curious patterns emerge on the away side. A double 
humped structure is seen: Soft hadrons correlated with a quenched jet have a 
distribution that is peaked at a finite angle away from the 
jet~\cite{star-cone,Adler:2005ee}, whereas they peak along 
the jet direction in vacuum. The variation of the peak 
with the centrality of the collision indicates that this is not due to 
the destructive interference of the LPM effect.
One proposal has been that the energy deposited by the away side 
jet excites a density wave in the medium which leads to a Mach cone 
like structure \cite{Casalderrey-Solana:2006sq,Ruppert:2005uz}. In 
these proceedings an alternate possibility will be explored.  

As pointed out in the introduction, the observed collective flow requires that the 
produced matter be strongly interacting~\cite{Gyulassy:2004zy}, with a very low 
viscosity~\cite{Teaney:2003kp}. There already exists a candidate model for such matter, 
consisting of a tower of 
colored bound states of heavy quasi-particulate quarks and gluons~\cite{Shuryak:2004tx}.
This model has however not fared well in comparison to lattice susceptibilities \cite{Koch:2005vg}, 
which rule out the possible existence of such bound states in the quark sector. Such 
calculations on the lattice do not, however, constrain the gluon sector of the plasma which 
may contain such states. It may indeed be possible for bound states to interact with 
gluons radiated by a hard parton passing through the dense matter.

The existence of coloured bound states in a deconfined plasma along with the assumption 
that these bound states have
excitations which may be induced by the soft gluon radiated from a jet, allows for 
a large index of refraction.
If the energy of the gluon is smaller than that of the first excited state, 
the scattering amplitude is attractive. As a result, the gluon dispersion relation
in this regime becomes space-like ($\ep > 1$) and Cherenkov radiation will occur \cite{Koch:2005sx,Dremin}.  
This is simply 
demonstrated in a $\Phi^3$ theory at finite temperature 
with three fields: $\phi$  a massless 
field representing the gluon and two massive fields $\Phi_1$ and $\Phi_2$ 
with masses $m_1$ and $m_2$ in a medium with a temperature $T$ 
(see Ref.\cite{Koch:2005sx} for details). 
The Lagrangian for such a system, ignoring the self-interactions between 
the scalars, has the form 

\bea
\mathcal{L} = \sum_{i=1}^2 \hf (\prt \phi_i)^2 + \hf (\prt \Phi)^2
+ \sum_{i=1}^2 \hf m_i^2 \phi_i^2 +  
g \Phi \phi_2 \phi_3.
\eea

\begin{figure}[htbp]
\hspace{3cm}
\resizebox{1.8in}{1.0in}{\includegraphics[0in,0in][6in,4in]{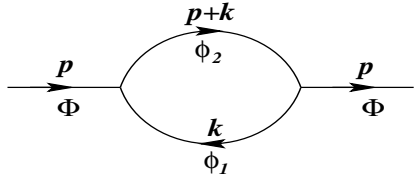}}
\caption{ General contribution to the self-energy 
of $\Phi$ due to transitional interaction. }
\label{fig2}
\end{figure}

The coupling constant $g$ is dimensionful; this along with all other 
scales will be expressed in units of the temperature. 
Ignoring issues related to vacuum renormalizability of such a theory, we 
focus on a study of the dispersion relation of the massless scalar in such an 
environment. The thermal propagator of $\Phi$ in the interacting 
theory is given in general as 
$D(p^0,\vp) =  [(p^0)^2 - |\vp|^2 - \Pi(p^0,\vp,T)]^{-1}, $
and the dispersion relation is given by the on-shell condition: 
$(p^0)^2 - |\vp|^2 - \Pi(p^0,\vp,T)=0$. Here, $\Pi(p^0,\vp,T)$ is the 
thermal self-energy of $\Phi$ due to loop diagrams such as the one 
shown in  Fig.\ref{fig2}. 

\begin{figure}[htbp]
\hspace{1cm}
\resizebox{1.75in}{1.75in}{\includegraphics[1.in,1.in][6.in,6in]{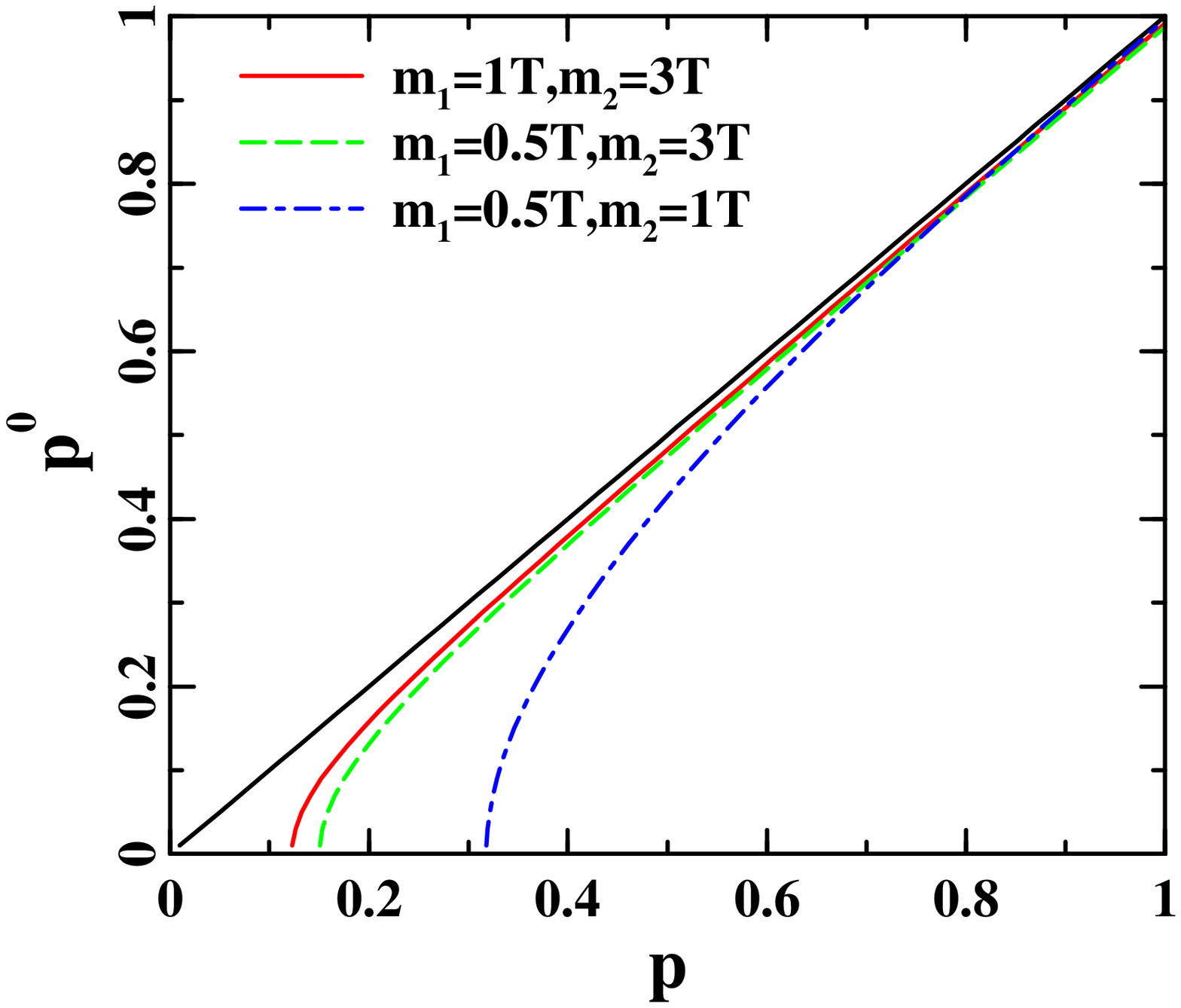}}
\hspace{2cm}
\resizebox{1.75in}{1.75in}{\includegraphics[1.in,1.in][6in,6in]{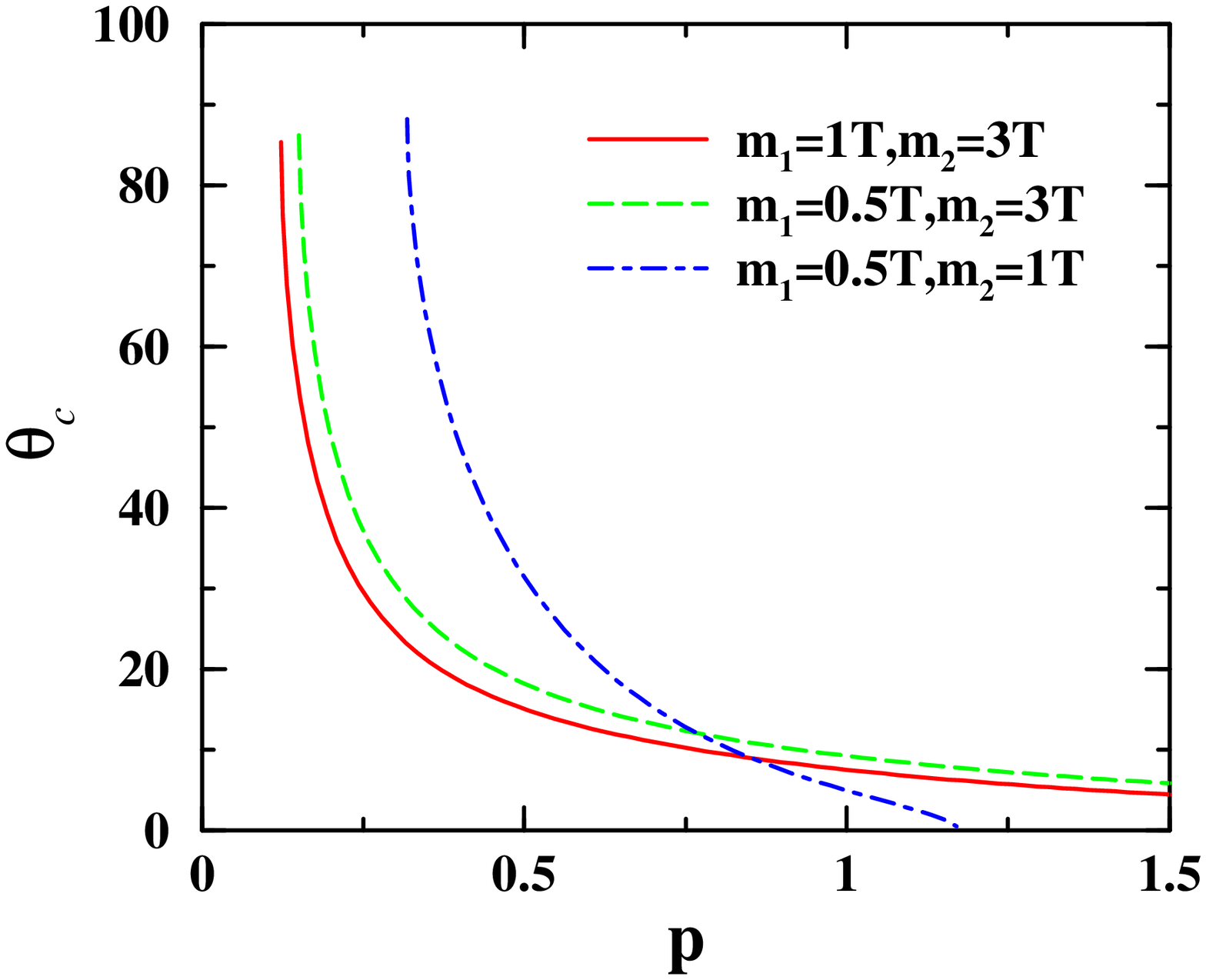}}
    \caption{ The left panel shows the dispersion relation of $\phi$ in a 
thermal medium with transitional coupling to two massive particle. The right panel 
shows the corresponding Cherenkov angles  versus the three momentum of the gluon.}
    \label{fig3}
\end{figure}

The resulting dispersion relations of the $\phi$ field 
for different choices of masses 
of $\Phi_1, \Phi_2$ are shown in Fig.~\ref{fig3} along with the 
corresponding Cherenkov angles of the radiation in the right panels.  
We obtain a space-like dispersion relation at low momentum which approaches 
the light-cone as the momentum of the gluon $(p^0,p)$ is increased. 
The variation of the corresponding angles may be actually detectable in 
current experiments. 
Even though we have studied 
a simple scalar theory, the attraction leading to Cherenkov-like 
bremsstrahlung has its origin in resonant scattering. Thus, the result 
is genuine and only depends on the masses of the bound
states and their excitations. Further experimental and theoretical investigations 
into such correlations may allow for a possible enumeration of the degrees of 
freedom in the produced excited matter. 

\section{Conclusions}

In these proceedings, the analysis of the properties of dense matter using the 
tool of jet correlations has been extended. Correlations between two 
high momentum hadrons within the same jet constitute an essential 
consistency check of the methodology and parameters of partonic energy loss. 
Correlations between a high momentum trigger and softer particles on the 
away side are shown to be more sensitive to the properties of the dense matter and
the prevalent degrees of freedom. 
Work supported by the U.S. Department of  Energy under grant nos. DE-FG02-05ER41367 and DE-AC03-76SF00098.

\end{document}